# Using Residual Dipolar Couplings from Two Alignment Media to Detect Structural Homology

Ryan Yandle[1], Rishi Mukhopadhyay[1], Homayoun Valafar[1]
[1]Department of Computer Science and Engineering, University of South Carolina, Columbia, SC, USA

**Abstract -** *The method of Probability Density Profile Analysis has been introduced previously as a tool to find a best match between a set of experimentally generated Residual Dipolar Couplings and a set of known protein structures. While it proved effective on small databases in identifying protein fold families, and for picking the best result from computational protein folding tool ROBETTA, for larger data sets, more data is required. Here, the method of 2-D Probability Density Profile Analysis is presented which incorporates paired RDC data from 2 alignment media for N-H vectors. The method was tested using synthetic RDC data generated with ±1 Hz error. The results show that the addition of information from a second alignment medium makes 2-D PDPA a much more effective tool which is able to identify a structure from a database of 600 protein fold family representatives.*

## 1 Introduction

One of the key restraints in computational protein folding is the dependency on a library of novel protein structures from which to base fold predictions. The deployment of the Structural Genomics Initiative offered an initial surge in finding these unique structures. More recently, however, the number of novel structures added to the Protein Data Bank (PDB) has been in decline (Figure 1). In 2007 there were only 4 new SCOP folds added to the data bank, down from 46 in 2006 and 78 in 2005 [1]. This steady decline in recent years suggests that the field needs a new procedure for rapidly identifying novel protein folds.

PDPA was first introduced in 2003 [2] as a method to rapidly find homologous structural candidates for a protein with unknown structure using minimal empirical data such as unassigned backbone RDCs. Previous work [3] was successful at identifying the correct structure for 1C99 as well as its nearest structural relatives from a pool of over 20 different structures. It was also demonstrated that PDPA can be used to confirm the most fit structural candidate for a protein fold given a list of 10 candidates supplied by ROBETTA [4]. These efforts were focused on using one dimension of data, generally a set of 15N-1H RDCs from a single alignment medium.

Even more recently, PDPA was expanded to explore the possibilities of using two dimensions (2-D) of data [5]. In this particular case 1-D PDPA was unable to pick out the correct structure for 16VPA [6] from a set of over 600 FSSP [7] family representatives. More information was needed in order to correctly identify 16VPA's structure as the top candidate. A first attempt at adding an additional channel of data was made by adding $C_\alpha$-$H_\alpha$ RDCs from the same alignment medium in conjunction with $^{15}$N-$^{1}$H data to construct a 2-D PDPA. For 16VPA, the use of 2-D data was able to give a more unique "finger print" for a particular structure.

This expansion of 2-D data can be pushed even further to incorporate data acquired from two alignment media. This will provide an even more unique view of the unknown structure in question, allowing for a more strict comparison to a library of known structures. Our research is focused on expanding the application of PDPA to multiple alignment media by using $^{15}$N-$^{1}$H RDCs from two alignment media.

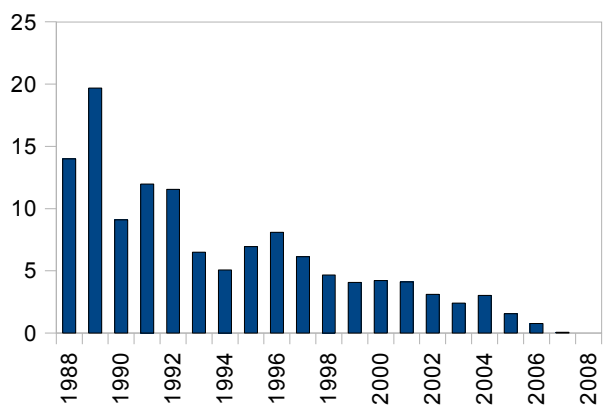

Figure 1: New SCOP folds as a percentage of new PDB submissions by year.

## 2 Data and Methodology

### 2.1 Residual Dipolar Couplings

A Residual Dipolar Coupling is a scalar measurement that can be interpreted as a constraint on the orientation of a particular bond between two spin ½ nuclei. This measurement can be established when two nuclei exhibit a partial alignment within the magnetic field of a NMR Spectrometer. RDCs were first observed in 1963, but the partial alignment of large biomolecular structures has not been possible until more recently [8]. Even more recently there have been many applications of this experimental data such as the refinement of computationally modeled protein structures [9]. The speed with which gathering unassigned RDC data can be completed makes using RDCs very favorable. Equation 1 presents a formal definition for calculating a RDC:

$$D_{ij} = D_{max} \cdot \langle \frac{3\cos^2(\theta)-1}{2} \rangle \quad (1)$$

Here $D_{ij}$ is the RDC measurement in hertz between two spin ½ nuclei in a large magnetic field. The angle between the intranuclear vector and the magnetic field is represented by θ, a time average is denoted by the angle brackets, and $D_{max}$ is the maximum observable value for a particular coupling, shown in Equation 2.

$$D_{max} = -\left(\frac{\mu_0}{4\pi}\right) \frac{\gamma_i \gamma_j \hbar}{2\pi^2 r_{ij}^3} \quad (2)$$

In this equation, μ0 is the magnetic permeability, γi and γj are the gyromagnetic ratios of our two corresponding nuclei, r is the fixed intranuclear vector and ℏ is Planck's constant.

For isotropically tumbling molecules, the observed time averaged value goes to zero thereby disallowing the observation of a RDC measurement. Therefore, the sample needs to be placed in an alignment medium such as liquid crystalline bicelles or filamentous bacteriophage. Dilution in such an alignment medium causes the molecules to tumble anisotropically, resulting in a non-zero time averaged value and allowing a RDC value to be observed.

$$D_{ij} = X^t \cdot S \cdot X \quad (3)$$

$$X = \begin{bmatrix} x_{ij} \\ y_{ij} \\ z_{ij} \end{bmatrix} \quad (4)$$

$$S = \begin{bmatrix} S_{xx} & S_{xy} & S_{xz} \\ S_{xy} & S_{yy} & S_{yz} \\ S_{xz} & S_{yz} & S_{zz} \end{bmatrix} \quad (5)$$

Manipulation of Equation 1 yields an equation for RDCs in matrix form (Equation 3). The unit vector between two nuclei is represented by X (Equation 4) and S is the symmetric and traceless Saupe [10] order tensor matrix (OTM) (Equation 5) that describes the anisotropic tumbling. Because of these properties, the OTM can be described with the five parameters $S_{xx}, S_{xy}, S_{xz}, S_{yy}, S_{yz}$. This OTM can be further decomposed to reveal the Euler rotation R and the matrix S' (Equation 6).

$$\begin{aligned} S &= R \cdot S' \cdot R^t \\ R &= R_z(\alpha) R_y(\beta) R_z(\gamma) \\ S' &= \begin{bmatrix} S'_{xx} & 0 & 0 \\ 0 & S'_{yy} & 0 \\ 0 & 0 & S'_{zz} \end{bmatrix} \end{aligned} \quad (6)$$

S' is composed of the principal order parameters $S'_{xx}, S'_{yy}, S'_{zz}$. These order parameters describe the strength of alignment of our molecule in the principal alignment frame, independent of any molecular frame. This equation allows us a second way to parameterize a given order tensor by the five parameters $S'_{xx}, S'_{yy}, \alpha, \beta, \gamma$.

It is clear that RDCs contain a wealth of structural orientation information. Assigned RDCs have been used for the assessment of secondary structure alignment and for protein fold family prediction [cite]. The costly step of acquiring RDCs, however, is assignment. Unassigned RDCs provide less information since their exact location within the sequence is unknown and therefore a particular RDC value is unable to be mapped to a specific chemical bond in the structure. Despite this fact, there is still enough information within an unassigned RDC data set to prove valuable.

## 2.2 1-D PDPA

1-D PDPA requires the gathering of a set of experimental unassigned RDCs and a library of structures for which the atomic coordinates are known. For instance, this library could be a set of fold family representatives from the latest SCOP listing. It is assumed that the RDCs contain some known level of noise. The RDCs are then analyzed through a Parzen density function in order to create a Probability Density Profile (PDP). In order to focus on the specific problem of structure matching, it is assumed that the principal order parameters have already been estimated through some other method such as maximum likelihood[11].

PDPA then takes this "query" PDP along with the estimated order tensor. A search is conducted over the set of 3 Euler rotations for every structure in the library. At each rotation, a PDP is calculated from RDCs of the library structure using the estimated order tensor. A scoring metric is then used to calculate the difference between the query PDP and the current library PDP of a structure at a particular Euler rotation. The best score for each structure is kept and all of the structures are sorted by their best score. The result is a ranking of "most fit" structures to our query.

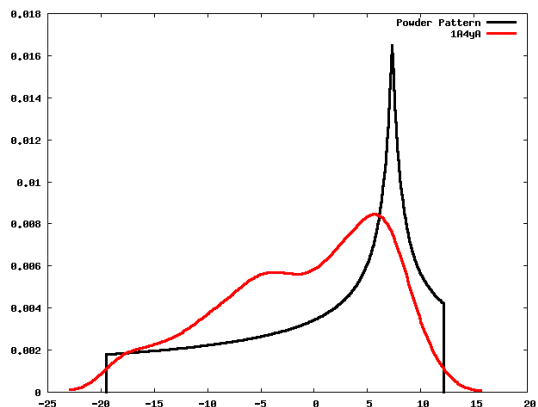

Figure 2: A PDP superimposed on the corresponding powder pattern.

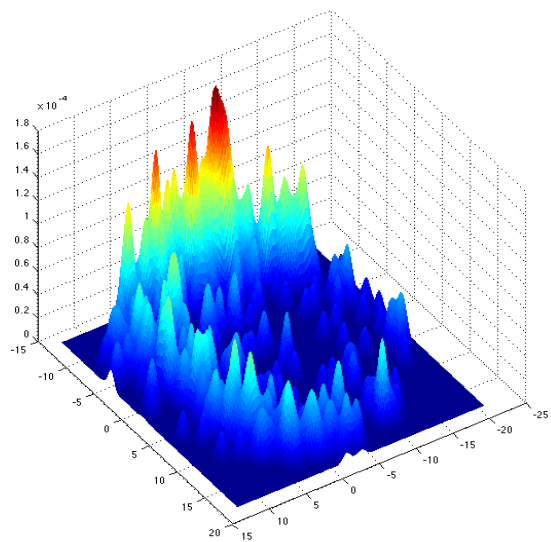

Figure 3: A 2-D Probability Density Profile for 1A4YA.

It can be shown that for a large enough sampling of 15N-1H backbone RDCs, the probability density function conforms to a powder pattern. Since real experimental samples do not yield this ideal number of RDC samplings, a typical PDP will have its own unique distribution (Figure 2).

An essential part to the statistical modeling of our profiles is the kernel function. 1-D PDPA uses a Parzen density function with a bandwidth appropriate to the estimated error. The Parzen density estimation function applied to 1-D RDCs in this fashion is defined here as a Probability Density Profile.

It was aptly noted in previous research that as the noise level of the RDCs is increased, the ability to properly distinguish a query PDP from its true best match fades.

As discussed in the introduction, 1-D PDPs have been shown to be insufficient in certain cases for determining structure similarity. It is critical, therefore, to introduce additional structural information to our profiles.

## 2.3  2-D PDPA

Transitioning from one alignment medium to two offers many advantages. With only one alignment medium, any given RDC value can be mapped to an infinite range of possible vectors. By adding the second alignment medium, the space of possible vectors for a given RDC pair contains no more than 8 vectors[3d]. This drastic reduction enables the probability density profile for a given protein to become much more unique.

2-D PDPA works much in the same way as its 1-D predecessor. The kernel estimation has been adjusted to produce PDPs that take into account both dimensions of data (Figure 3). A standard distance metric to determine PDP similarity is used.

At first blush, the utilization of two alignment media in PDPA would seem computationally infeasible since the addition of a second alignment media would require a search over another set of 3 Euler angles. At a resolution of 5° rotations, this would require a search of 2.5657e+09 (37$^6$) possible configurations for each candidate in order to describe its alignment in two independent media. Recent work, however, has alleviated the need for such a search.

## 2.4  2-D Order Tensor Estimation

The work done in [12] has produced a method of quickly obtaining an estimated set of relative order tensors for RDC readings from two alignment media. For the rest of this discussion, the order tensor for alignment 1 will be referred to as S1 and the second alignments tensor will be denoted S2.

After 2-D OTM estimation, $S_1$ is given in the principal alignment frame. This is to say that S1 is of the diagonal form shown in (Equation 6) and $S_2$ is composed of the 5 order parameters from Equation 5. The curious reader is referred to [12] for more information on 2-D OTM estimation.

Since the 2-D OTM estimation renders two alignment tensors relative to each other, the search is once again limited to a single set of 3 Euler rotations. Because the 4-fold symmetry inherent in RDCs is eliminated when using two or more alignment media, the search must be extended to take into account not only the original starting position, but three other possible starting potions that are 180° rotations about each axis. Therefore, at a search resolution of 5° rotations, there are 202,612 (4 x 37³) configurations for each library protein that must be analyzed. This is a much more realistic search space compared to the previous estimate.

## 2.5 Parallelization

PDPA intrinsically exhibits strong data parallelism. Given that the search space for 2-D PDPA is 4 times larger than 1-D, parallelization is crucial to achieving an acceptable run time. A serial version of 2-D PDPA could take up to 7.5 days to complete with a library of 600 structures. The parallel version of 2-D PDPA, which runs on a 64 node cluster with dual CPU 2.0ghz dual core processors, can complete the same task in only two hours yielding a speedup of 90x. It should be noted that although this is a tremendous increase in speed, the maximum speedup that could be achieved with a larger scale HPC is much higher. There are also many more optimization techniques that could be applied to 2-D PDPA such as pre-processing library filtration based on protein size.

|   | $S_{xx}$ | $S_{xy}$ | $S_{xz}$ | $S_{yy}$ | $S_{yz}$ |
|---|---|---|---|---|---|
| $S_1$ | 3.000E-04 | 0.000E-04 | 0.000E-04 | 5.000E-04 | 0.000E-04 |
| $S_2$ | 1.066E-04 | 2.367E-04 | 3.603E-04 | -1.464E-04 | 4,323E-04 |

Table 1: The order tensors used to simulate RDC data

## 2.6 Experimental Setup

The experiment consists of four proteins of varying size; 1SF0 (77), 1IOM (154), 16VPA (366) and 1A4YA (460). The library of known structures is made up of 600 fold family representatives taken from FSSP 2003 [cite], supplemented with the four query proteins listed above. For each protein, $^{15}$N-$^{1}$H backbone RDCs are calculated with ±1hz error in REDCAT [13] using the order tensors in table 1.

Next, the synthetic data is used to estimate two relative order tensors and generate the query PDP. These estimated tensors, along with the query PDP, are then input into 2-D PDPA. The result is an ordered listing of each library structure and its best score. We have filtered this output based on protein size. In the results we only included proteins whose size is within ±20% of the query protein's size.

| | | |
|---|---|---|
| 1SF0 | 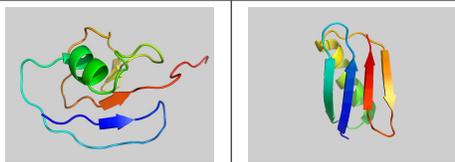 | |
| 1IOM | 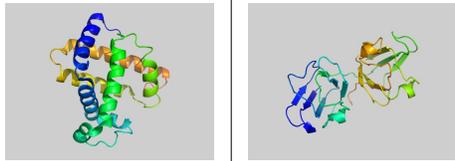 | |
| 16VPA | 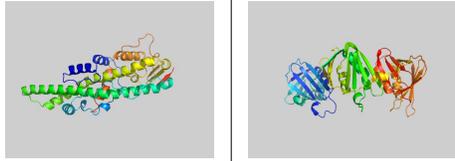 | |
| 1A4YA | 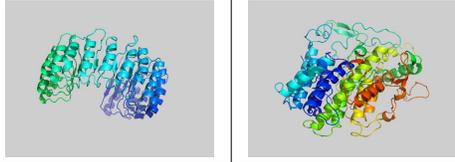 | |

Table 2: The top two structures for each run of 2-D PDPA

## 3 Results

For this experiment, a manhattan distance metric was used to compute the score for each candidate. Hence, a lower score means more similarity between the query and library PDPs. For three out of the four runs, 2-D PDPA identified the correct structure as first. In addition to the correct structure being first, all top scores have a significantly lower score compared to the next four, making each top candidate a clearly distinguishable leader.

The results for 1SF0 were not as good. 1SF0 was the smallest protein in the group of experimental runs. Because of this, there were significantly fewer observed RDC values. This in turn made the PDP less defined. The lack of RDCs also affected the relative order tensor estimations. These estimations were farther off from the original order tensors used. When 1SF0 was rerun with the original order tensors, the correct structure was first by a significant lead. The effects of a poor 2-D relative order tensor estimation need to be explored further.

| 1SF0* | | 1IOM | | 16VPA | | 1A4YA | |
|---|---|---|---|---|---|---|---|
| 1SF0 | 0.88835 | 1IOM | 0.80910 | 16VPA | 0.69469 | 1A4YA | 0.60190 |
| 1P7E | 1.18836 | 1AMM | 1.03750 | 2POLA | 0.82501 | 1VNS | 0.69852 |
| 1IIEA | 1.20415 | 1EX2A | 1.04411 | 6PRCC | 0.84388 | 1DERA | 0.70791 |
| 1FAFA | 1.20779 | 1BV1 | 1.04419 | 1BT3A | 0.84616 | 1H2RL | 0.71315 |
| 1KJS | 1.2312 | 1G3P | 1.04818 | 1G71A | 0.84910 | 1QDBA | 0.71315 |

Table 3: The results for 2-D PDPA with four proteins of varying size and structure. Each candidate has the top 5 scores listed below it.
*For 1SF0, the results displayed were run with the real order tensor.

## 4  Conclusion

2-D PDPA has a proven capability of identifying similar structures with the presence of error and estimated order tensors. The addition of a second alignment medium, which was not computationally feasible until recently, has allowed each structure to become more uniquely expressed in their PDP. This new pipeline of information opens up the possibility of utilizing experimental data consisting of RDCs from multiple alignment media to correctly identify fold families.

It is envisioned that during target selection, unassigned RDCs can be quickly gathered and run through PDPA in order to determine whether or not the target protein would be a good candidate to study further. If the target is indeed a good candidate, then more effort can be put into determining the structure of the protein. Afterwards, the RDCs that were initially collected can be used in PDPA to confirm the proposed structure. In the case that the target does not seem to be a good candidate, then the researcher has only spent about a week's worth of time collecting unassigned RDCs and can easily move on to the next proposed target. This allows for the initial RDC collection to be useful for either case of target selection.

## 5  Future Work

There are still many aspects of 2-D PDPA that have yet to be explored. The possibility of a more favorable distance metric exists. The standard block distance is yielding good initial results, but a more in depth analysis of other metrics is needed. The parallelization of PDPA offered a tremendous improvement in running time. There still exist many optimization techniques such as filtration by protein size that can be applied in order to speed up PDPA. A careful examination of the use of real world experimental data is also needed in order to fully exhibit the robustness of the system. We would like future versions of PDPA to use a much larger library, like the PDB [14].

In order to use the entire PDB and still have a reasonable running time, we propose first using a library of fold representatives from the latest SCOP or CATH [15] releases. PDPA would then take the top 5 fold families and re-run with a library consisting of all the sub-family representatives.

The ultimate goal of our work is to release PDPA as a web application. This application could incorporate many of the tools from our research group such as REDCAT, TALI [16] and 2-D order tensor estimation. Researchers without access to a high performance computer could easily examine their experimental data. This would provide a truly robust mechanism with which to extract structural information from a minimum set of residual dipolar couplings.

## 6  Acknowledgements

Funding for this project was provided by NSF Career Grant # MCB-0644195 awarded to Dr. Homayoun Valafar.

We would like to thank the University of South Carolina's High Performance Computing Group for the computing time used in this research.